\documentclass[11pt,a4paper]{article}
\usepackage{graphics}
\usepackage{epsfig}
\usepackage{fancyhdr}
\usepackage{amssymb}
\usepackage{amsmath}

\usepackage[T1]{fontenc}
\usepackage[utf8]{inputenc}
\usepackage{authblk}

 \title{The black disk and the dip in the differential elastic cross section at asymptotic energy}

\author[1,2]{Irais Bautista\thanks{irais@fpaxp1.usc.es}}
\author[1]{Jorge Dias de Deus\thanks{jorge.dias.de.deus@ist.utl.pt}}

\affil[1]{CENTRA, Instituto Superior T\'ecnico, Universidade T\'ecnica de Lisboa, \\ Av. Rovisco Pais, P-1049-001 Lisboa, Portugal}
\affil[2]{IGFAE and Departamento de F\'isica de Part\'iculas, Univ. of Santiago de Compostela, 15782, Santiago de Compostela, Spain}

\begin{document}

\maketitle

\begin{abstract}
We test the validity of the black disk limit in elastic scattering by studying the evolution of the dip in the scaling variable $\tau=-t_{D}\sigma^{tot}$, where $t_{D}$ is the transverse 
momentum squared at the dip and $\sigma_{tot}$ the total cross section. As $s\rightarrow \infty $ and $-t_{D} \rightarrow 0 $, $\tau$ may consistently be approaching the black disc value,
$\tau \xrightarrow[ \sqrt{s}\rightarrow \infty ]{} \tau_{BD}=35.92$ GeV$^{2}$ mb.

\end{abstract}

Recent results from LHC (pp scattering at 7 TeV) and from Auger Observatory (pAir at 57 TeV) on  total and elastic cross sections, [1-3], may be quite relevant to improve our understanding of the asymptotic behavior, $\sqrt{s} \rightarrow \infty$, of cross-sections.

There are two important theorems obtained by making use of fundamental concepts as analyticity, crossing symmetry and unitarity:
\vspace{5mm}

1) Froissart bound [4],

\begin{equation}
\sigma(s)^{tot} \sim 2 \pi R^{2}(s) \sim log^{2}(s/s_{0}).
\end{equation}

The proof of the theorem requires the existence of a maximum angular momentum $L(s)$, proportional to some radius $R(s)$, above which the contributions to the partial wave sum are negligible.
\vspace{5mm}

2) Geometric Scaling GS [5,6,7]
\vspace{10mm}
In the limit of Froissart behavior, (1), it follows that 
\begin{equation}
ImF(s,t)=ImF(s,0) \varphi(\tau),
\end{equation} 
where $Im F(s,t)$ is the imaginary part of the amplitude 
and $\varphi$ an entire function of the scaling variable $\tau$,

\begin{equation}
\tau \equiv -t \sigma^{tot}.
\end{equation}

One should notice that $t$ and the impact parameter $b$ are conjugate variables with the result that $\beta$,
\begin{equation}
\beta^{2}= b^{2}/\sigma^{tot},
\end{equation}
is also a scaling variable.
GS ideas and phenomenology were developed in [6] and [7].

One should also notice that the original GS does not agree with data. The ratio $\sigma^{el}/\sigma^{tot}$ is predicted to be constant while a clear growth with energy is seen in data [8].

In order to see why is it so, let us write (see, for instance, [9]):

\begin{equation}
\sigma^{tot}(s)=2 \pi \int db^{2} Im G(s,b) \xrightarrow[ GS] {} 2 \pi R^{2}(s) \int_{0}^{1} d\beta^{2} Im G(\beta)
\end{equation}
and 
\begin{equation}
\sigma^{el}(s)=\pi \int d b^{2} [Im G(s,b)]^{2} \xrightarrow[ GS] {} \pi R^{2}(s) \int_{0}^{1} d \beta^{2} [Im G(\beta)]^{2},
\end{equation}
 where $G(s,b)$ is the elastic amplitude and in (6) the real part  was neglected. From (5) and (6) one immediately sees that $\sigma^{el}/\sigma^{tot}= const \leq 1/2$.
 
 The relevant cross- sections, (5) and (6), contain explicit dependence on energy via $R^{2}(s)$, the quantity controlling the size and range of the interactions. But energy should also affect the quark- gluon matter density, showing evolution towards saturation. We introduce a second function depending on energy, $f(s)$, to describe evolution of matter density.
 
 We shall next make a grey disk approximation, and identify the averaged in $\beta$ of $Im G(\beta)$ with $f(s)$;
 \begin{equation}
 <Im G(\beta)> \simeq f(s),
 \end{equation}  
 with 
 \begin{equation}
 \frac{df}{ds} \geq 0,
\end{equation} 
and 
\begin{equation}
f(s) \xrightarrow[s \rightarrow \infty] {} 1.
\end{equation}
Equation (9) is a consequence of unitarity saturation in the black disk limit. 
Note that (8) says that blackness increases with energy.

Making use of (5), (6) and (7) we obtain:

\begin{equation}
\sigma^{el}(s)/ \sigma^{tot}(s)=\frac{1}{2} f(s),
\end{equation}
in violation of GS, and 
\begin{equation}
\sigma^{tot}(s)= 2 \pi R^{2}(s) f(s).
\end{equation}
Asymptotically, (10) and (11) satisfy GS.
 Note that the function $f(s)$, describing unitarity saturation, was introduced in [11]:

\begin{equation}
 f(s)=1-e^{-\bar{\Omega} (s)},
 \end{equation}
 the opacity  $\bar{\Omega} (s)=2(\gamma_{1}+\gamma_{2} ln(s) + \gamma_{3} ln^{2} (s))$, and $\gamma_{1}=0.29$,
  $\gamma_{2}=0.0191$, $\gamma_{3}=0.0013352$ to keep common notations with [10]. 
In both cases, (10) and (11), the asymptotic behavior, as energy increases, is reached from below (see [12]).

The physics of (10) and (11), in the $f(s)\rightarrow 1$ limit, is black disk physics. In (10) we obtain
\begin{equation}
\sigma^{el}(s)/\sigma^{tot}(s) \rightarrow \frac{1}{2}.
\end{equation}
In (11), having in mind that for the black disk $B(s,t=0) \rightarrow R^{2}/4$, where $B$ is the slope parameter we arrive at 
\begin{equation}
\sigma^{tot}/B(s,0) \rightarrow 8 \pi
\end{equation}
Relations (13) and (14), see [13], are well known black disk relations.

Making use of the analytical properties of amplitudes and cross- sections it was possible to estimate the ratio $\sigma^{inel}(s)/\sigma^{tot}(s)$ at asymptotic energies to obtain a value $(0.509 \pm 0.011)$ [14], consistent with the naive expectation for a black disk (see [15] for general discussion). Our neglect of $Re G(\beta)$, in particular in the forward peak, is a way of having, asymptotically, the black disk.

We turn next to GS and write, see (2) and (3)
\begin{equation}
\frac{d\sigma}{dt}(t)/\frac{d\sigma}{dt}(0) \xrightarrow[\sqrt{s}\rightarrow \infty]{} \varphi^{2}(\tau),
\end{equation}
where $\tau$ is the scaling variable. If GS was exact (13) would be exact. If it is just true asymptotically we have to concentrate in the limit $\sqrt{s}\rightarrow \infty$.

In order to test GS let us consider the evolution of the position of the minimum $\tau_{D}=-t_{D} \sigma^{tot}$, seen in the range $\sim$ (20 GeV $\leq \sqrt{s} \leq 7 $ TeV). One observes that
 $\sigma^{tot}$ increases with energy (see (10), $\sigma^{tot} \sim R^{2}(s) f(s)$) so one needs $-t_{D}$ to decrease with energy. As we do not have a strict prediction for the evolution of $-t_{D}$
we write, for instance, $-t_{D}\sim \frac{1}{\sigma^{tot}}$ and we have GS for any value of $\sqrt{s}$.

As GS can only be correct asymptotically we write
\begin{equation}
-t_{D}=\frac{1}{2\pi R^{2}(s)} \frac{1}{f(s)^{\alpha}} \tau_{BD}
\end{equation}
with $\sigma^{tot}(s)$, given by (11), and  $\tau_{BD}=35.92$ GeV$^{2}$ mb [16] being the black disk $\tau$ and $\alpha$ is a parameter. If $\alpha=1$ GS works at all energies. Experimentally we obtained $\alpha=1.47$, and GS is asymptotic.

In Fig.(1) 
we present the obtained energy dependence of 
$-t_{D}$, (16). In Fig. (2) we show $\tau=-t_{D} \sigma^{tot}$ as a function of $\sqrt{s}$. $\tau$ seems to approach $\tau_{BD}=35.92$ GeV$^{2}$ mb, in a slow process. The star (*) in Fig. (2) corresponds to our expectation for $\sqrt{s}=14$ TeV, using information from Fig. (1) At the star $\tau_{*}=44.9$ ($\sqrt{s}=14$ TeV).
In conclusion, we find at present LHC energies indications that we are approaching black disk behavior $(\sigma^{el}/\sigma^{tot}) \rightarrow 1/2$, $\sigma^{tot}/B(s) \rightarrow 8 \pi$, and $\tau_{DIP} \rightarrow \tau_{BD} =35.92$ GeV$^{2}$ mb. However we are still far from asymptopia.

\begin{figure}
\begin{center}
    \begin{tabular}{cc}

     \resizebox{75mm}{!}{\includegraphics{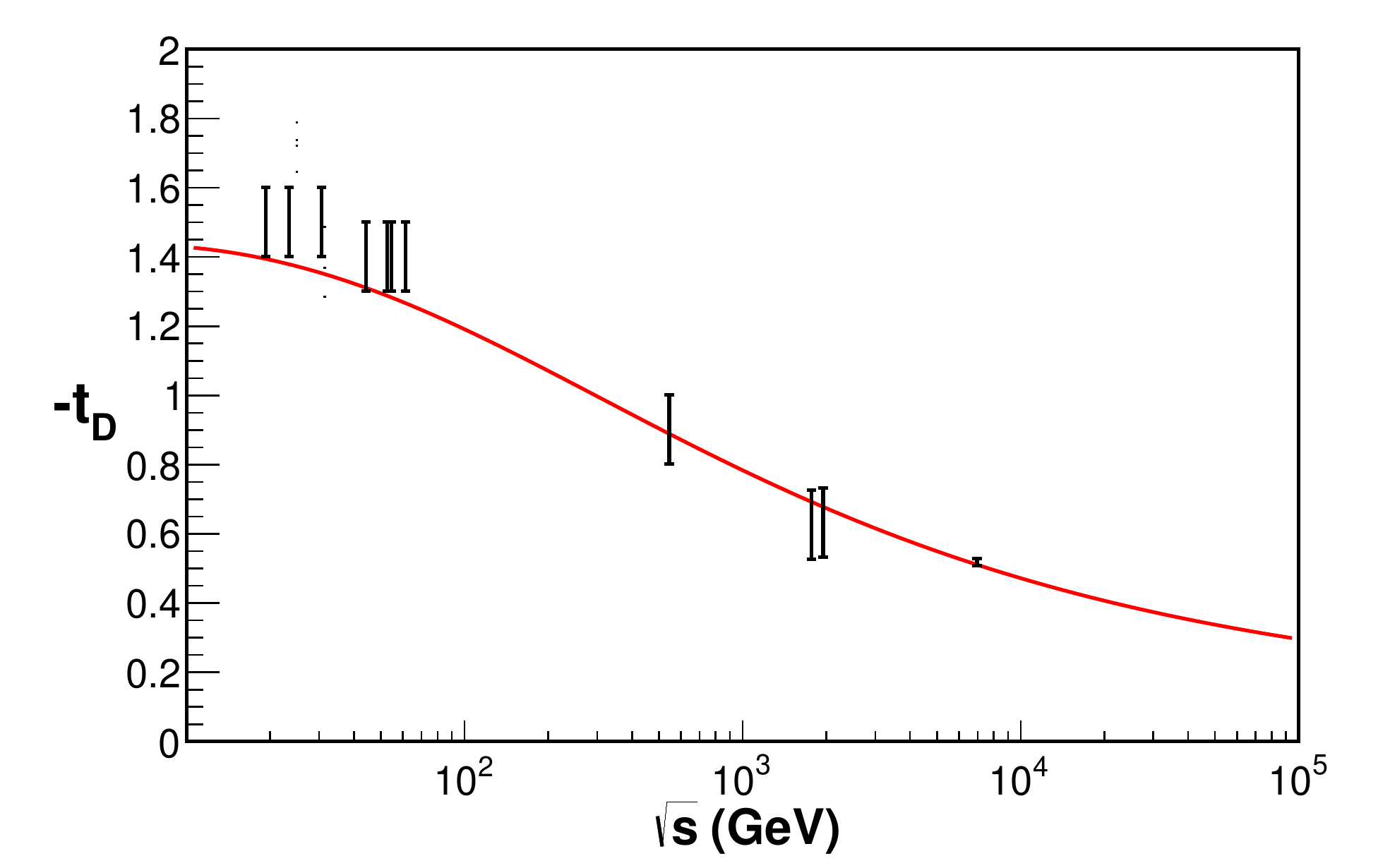}}\\
     \mbox{a)} & \mbox{b)}
\end{tabular}
\caption{  
$-t_{D}$  as a function of energy $\sqrt{s}$.  Here $R(s)$ is given by the parametrization $R(s)=R_{0} ln(s/s_{0})$, with $R_{0}=.0936$ mb$^{1/2}$, and $\sqrt{s_{0}}=2.216*10^{-9}$ GeV , (16).}
\end{center}     

\end{figure}

\begin{figure}
\begin{center}
     \resizebox{100mm}{!}{\includegraphics{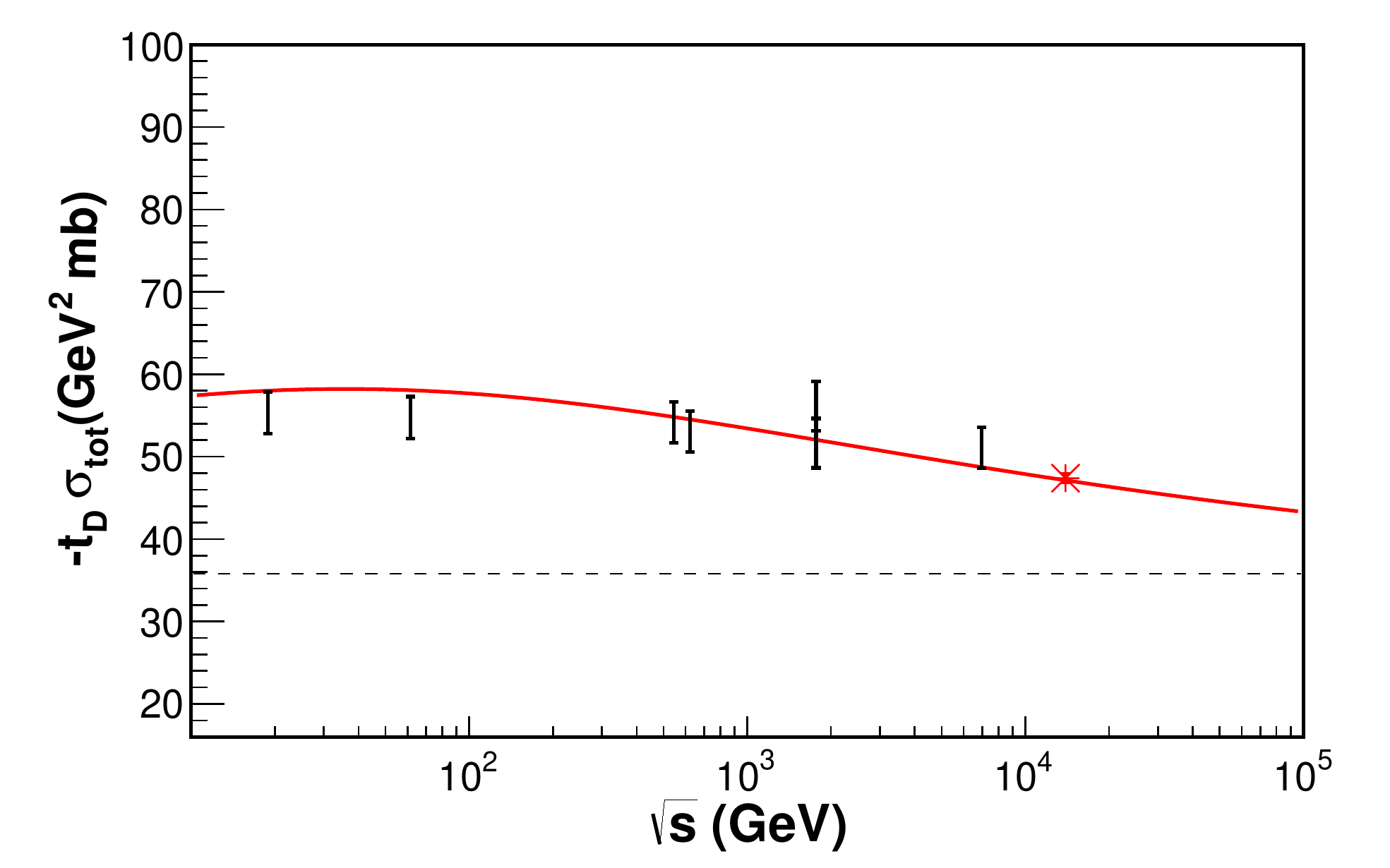}}\\ 
        \caption{Solid line shows $-t_{D} \sigma_{tot}$ as function of energy from (11) and $t_{D}$ as in previous figure. The dashed line shows the black disk limit. The star corresponds to expectation for $\sqrt{s}=14$ TeV.}
    \end{center}     
\end{figure}

\section*{Acknowledgments}
We would like to thank Pedro Brogueira who helped in characterizing the features of the black disk. We thank Daniel Fagundes and Marcio Menon for discussions. JDD acknowledge the support of Funda\c c\~ao para a Ci\^encia e a Tecnologia (Portugal) under project CERN/FP/116379/2010. 
IB was supported by SFRH/BD/51370/2011 from Funda\c c\~ao para a Ci\^encia e a Tecnologia (Portugal).

\end{document}